\documentclass[conference]{IEEEtran}
\IEEEoverridecommandlockouts
\usepackage{cite}
\usepackage{amsmath,amssymb,amsfonts}
\usepackage{algorithmic}
\usepackage{graphicx}
\usepackage{textcomp}
\usepackage{xcolor}
\usepackage{afterpage}
\usepackage{float}
\usepackage{subcaption}
\usepackage{caption}
\usepackage{array}

\def\BibTeX{{\rm B\kern-.05em{\sc i\kern-.025em b}\kern-.08em
    T\kern-.1667em\lower.7ex\hbox{E}\kern-.125emX}}
\begin{document}

\title{Inter-Frame Coding for Dynamic Meshes via Coarse-to-Fine Anchor Mesh Generation\\
% \thanks{}
}

\author{
    He Huang$^{\star}$ \qquad Lizhi Hou$^{\star}$ \qquad Qi Yang$^{\dagger}$ \qquad Yiling Xu$^{\star}$\\
    \\
    $^{\star}$ Shanghai Jiao Tong University, $^{\dagger}$ University of Missouri-Kansas City \\
    \{huanghe0429, lizhi.hou, yl.xu\}@sjtu.edu.cn, littlleempty@gmail.com
}

\maketitle

\captionsetup[subfigure]{labelformat=simple, labelsep=space}

\setlength{\abovedisplayskip}{2.4pt} % Adjust the space above the equation
\setlength{\belowdisplayskip}{2.4pt} % Adjust the space below the equation
    
\begin{abstract}
In the current Video-based Dynamic Mesh Coding (V-DMC) standard, inter-frame coding is restricted to mesh frames with constant topology. Consequently, temporal redundancy is not fully leveraged, resulting in suboptimal compression efficacy. To address this limitation, this paper introduces a novel coarse-to-fine scheme to generate anchor meshes for frames with time-varying topology. Initially, we generate a coarse anchor mesh using an octree-based nearest neighbor search. Motion estimation compensates for regions with significant motion changes during this process. However, the quality of the coarse mesh is low due to its suboptimal vertices. To enhance details, the fine anchor mesh is further optimized using the Quadric Error Metrics (QEM) algorithm to calculate more precise anchor points. The inter-frame anchor mesh generated herein retains the connectivity of the reference base mesh, while concurrently preserving superior quality. Experimental results show that our method achieves 7.2\% \( \sim \) 10.3\% BD-rate gain compared to the existing V-DMC test model version 7.
\end{abstract}


\addtolength{\belowcaptionskip}{-6pt}
\afterpage{
\begin{figure}[t]
    \centering
    \includegraphics[width=\columnwidth]{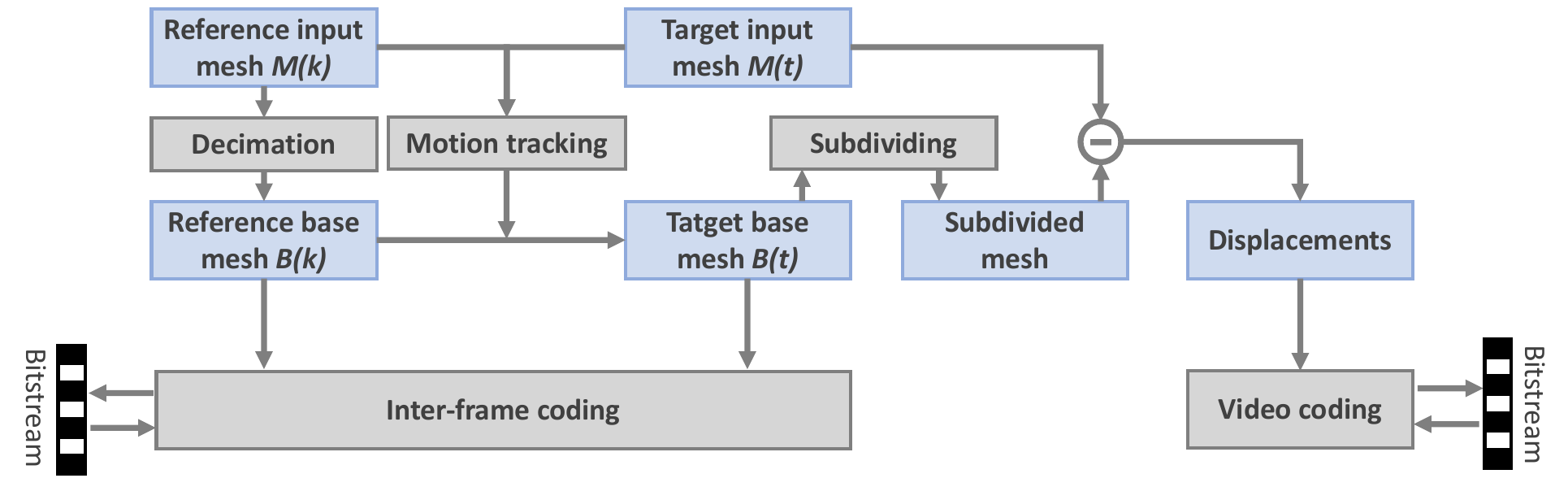}
    \caption{Architecture of inter-frame coding in V-DMC for input meshes with constant topology.}
    \label{inter-frame coding framework}
\end{figure}
\begin{figure}[t]
    \centering
    \includegraphics[width=0.95\columnwidth]{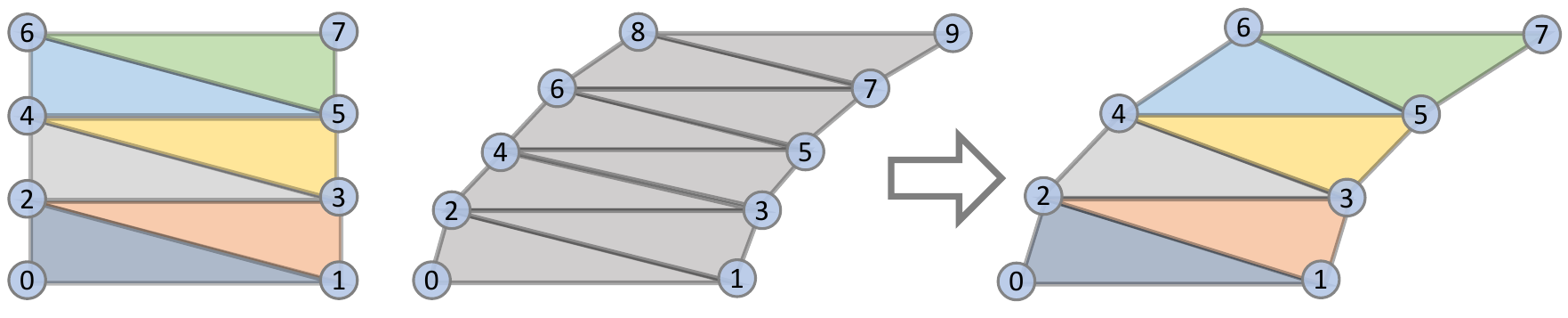}
    \caption{Effects of anchor mesh generation. From left to right:
reference base mesh (left), target input mesh (middle), and anchor mesh (right). The anchor mesh has a one-to-one vertex and face correspondence with the reference base mesh.}
    \label{topology}
\vspace{-10pt}
\end{figure}
}

\section{Introduction}
With rapid advancements in 3D capture, modeling, and rendering, 3D immersive content is increasingly being applied in Augmented Reality (AR), Virtual Reality (VR), and Autonomous Driving (AD). In these applications, 3D meshes are the most prevalent media format because they are well-suited for GPU rendering and highly applicable to interactive and real-time 3D tasks. A dynamic mesh sequence consists of consecutive static mesh frames, each of which has a ground of polygonal faces. Compared with traditional 2D media, 3D mesh requires substantial data for vivid 3D representation, highlighting the need for efficient compression methods.

Although existing mesh compression technologies, such as Draco\cite{rossignac1999edgebreaker} and TFAN\cite{mamou2009tfan}, have demonstrated superior performance in static meshes, they are inadequate for handling dynamic mesh sequences. To address this challenge, the 3D Graphics and Haptics Coding (3DG) of the Moving Picture Experts Group (MPEG) has formed a standard framework for video-based dynamic mesh coding(V-DMC) to explore more effective coding algorithms \cite{mammou2022ideo}.

For intra-frame coding, V-DMC decomposes an original mesh into two key components for compression: a decimated base mesh and a detailed displacement field to efficiently compress the geometry information. The base mesh preserves the fundamental properties of the input mesh while reducing the number of vertices and faces. The differences between the input mesh and the subdivided base mesh are calculated and encoded as displacements.

Fig.~\ref{inter-frame coding framework} illustrates the workflow of V-DMC inter-frame coding for dynamic mesh sequences with constant topology. For the target mesh frame \textit{M(t)}, V-DMC estimates a motion field by tracking the corresponding vertices between \textit{M(t)} and the reference mesh frame \textit{M(k)}. Subsequently, the reference base mesh \textit{B(k)}, which is decimated from \textit{M(k)}, is deformed into the current base mesh \textit{B(t)} using the estimated motion field. Therefore, \textit{B(t)} has the same connectivity with \textit{B(k)}. After subdividing \textit{B(t)}, the residuals between the nearest point on the surface of \textit{M}(t) and each vertex of the subdivided base mesh are calculated as displacements. This method can significantly reduce data volume by using a shared base mesh.  However, for mesh sequences that exhibit a time-varying topology, V-DMC cannot compute the motion field by tracking the corresponding vertices to generate the target base mesh.

While dynamic point cloud compression technologies \cite{zhu2020view, fan2023multiscale, xia2023learning} have reached a high degree of maturity, the domain of dynamic mesh compression remains underexplored. In the pioneering research on dynamic mesh coding \cite{habe2004skin,projection,Static}, meshes are projected onto 2D images, which are subsequently encoded using video codecs. These methods cannot efficiently encode the connectivity between vertices because of the projection error on 2D images. Other methods attempt to utilize transformations to represent dynamic meshes \cite{han2007time, yamasaki2010patch, jin2023inter}. However, they are limited to geometry compression and require constant connectivity.

% Other methods such as projection \cite{habe2004skin, projection} and transformation \cite{han2007time, yamasaki2010patch, jin2023inter} are limited to geometry compression and require constant connectivity. Therefore, these methods are also unsuitable for meshes with temporal inconsistencies.

To address the issue, we propose a novel inter-frame coding scheme for meshes with time-varying topology. Specifically, we utilize a coarse-to-fine method to generate a high-quality anchor mesh that shares the same topology as the reference base mesh. Initially, we generate a coarse anchor mesh through an octree-based nearest neighbor search \cite{octree} and motion estimation between the reference base mesh and the target input mesh. The rapidly generated anchor mesh ensures the time consistency of the topology. Furthermore, the fine anchor mesh is optimized using the Quadric Error Metrics (QEM) algorithm \cite{garland1997surface} based on the coarse mesh to enhance its quality, resulting in the high-quality anchor mesh. One toy example of this coarse-to-fine method is depicted in Fig.~\ref{topology}. Additionally, we have developed a more precise method for adaptive displacement quantization that takes into account the significance of each vertex. The significance is calculated based on its neighbor count. The experimental results show that our proposed method can achieve 7.2\% \( \sim \) 10.3\% BD-rate gain without a significant increase in encoding and decoding time. The ablation study further demonstrates that each constituent module confers distinct advantages, in which the QEM-based refinement shows a particularly pronounced efficacy.

\addtolength{\belowcaptionskip}{-6pt}
\begin{figure}[t]
    \centering
    \includegraphics[width=0.95\columnwidth]{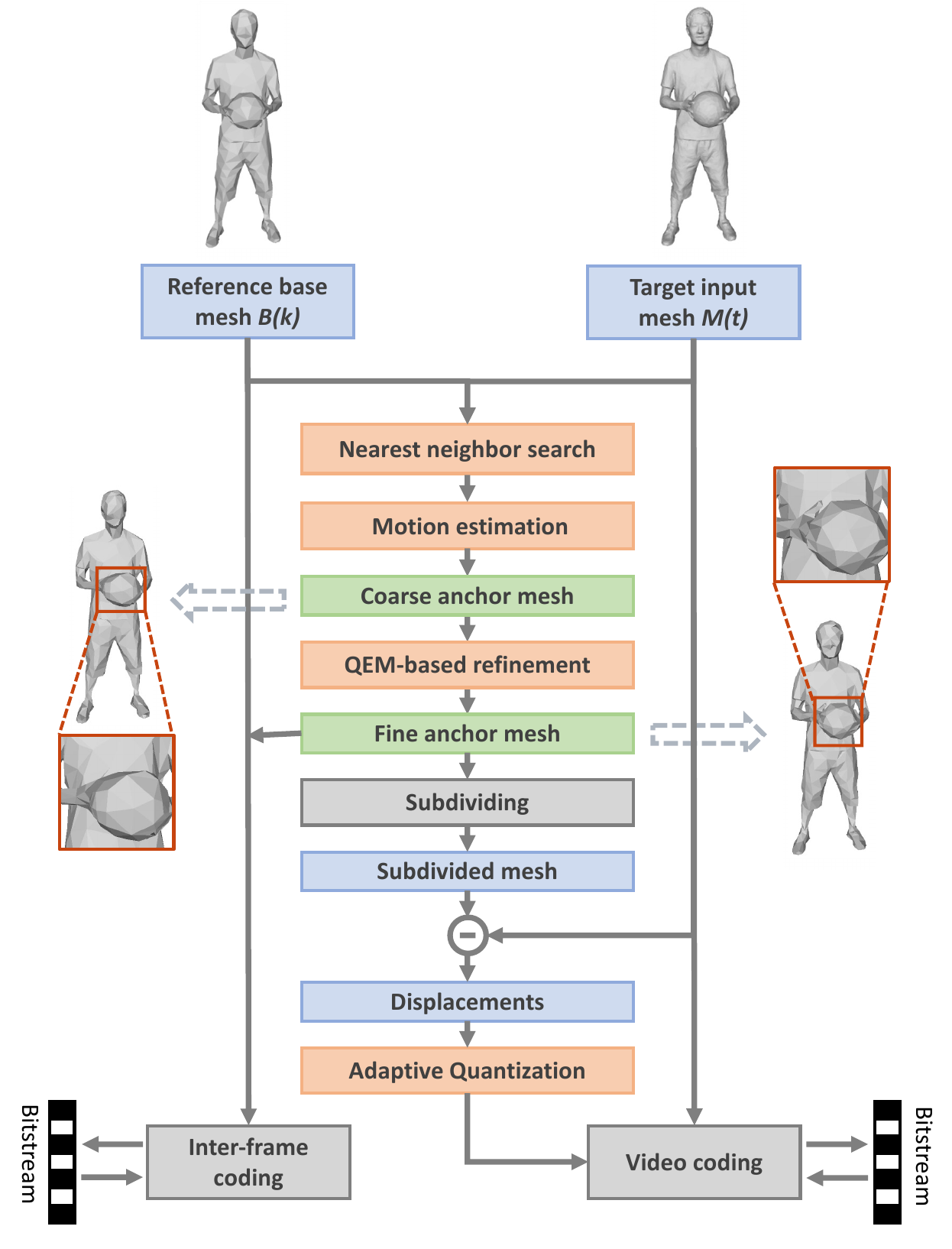}
    \caption{Architecture of proposed coarse-to-fine inter-frame anchor mesh generation method.}
    \label{QEM matched mesh framework-pdf}
\vspace{-10pt}
\end{figure}

\section{PROPOSED METHODS}
In this section, we detail our proposed coarse-to-fine inter-frame anchor mesh generation method. As depicted in Fig.\ref{QEM matched mesh framework-pdf}, we utilize \textit{B(k)} and \textit{M(t)} to generate the anchor mesh. Initially, for every vertex in \textit{B(k)}, the algorithm identifies a corresponding point in \textit{M(t)} using an octree-based nearest neighbor search and motion estimation to generate the coarse anchor mesh. Subsequently, the algorithm evaluates and selects the edge in \textit{M(t)} with the smallest metric for collapse around the corresponding point. The collapsed point serves as the fine anchor vertex, establishing a one-to-one correspondence with the vertex in \textit{B(k)}. Moreover, the displacements computed from \textit{M(t)} and the fine anchor mesh are adaptively quantized.

\subsection{Coarse-Stage Anchor Mesh Generation}
The main issue for frames with time-varying topology is that \textit{B(t)} often cannot maintain the same topology as \textit{B(k)}. To address this, we propose using the octree-based nearest neighbor search method, which can quickly generate a mesh with the same connectivity. Specifically, for each vertex in \textit{B(k)}, we build an octree and use the nearest neighbor search in \textit{M(t)} to find the corresponding nearest vertex. By employing the aforementioned vertex matching technique, we reformulate the challenge of time-varying topology into a geometric matching problem. Therefore, the objective is to identify a set of vertices $V_{t}$ in \textit{M(t)} such that the aggregate Euclidean distances between these vertices and those in \textit{B(k)} are minimized, as demonstrated in:
\begin{equation}
\mathrm{arg}\min_{V_{t}^{j}} \sum_{i} \left |{Dist} (V_{k}^{i}, V_{t}^{j}) \right | \label{min}
\end{equation}
\begin{equation}
Dist(A,B) = \sqrt{(A-B)^{2}} \label{dist}
\end{equation}
where $V_{k}^{i}$ is the vertex of \textit{B(k)} and $V_{t}^{j}$ is the corresponding vertex of \textit{M(t)}. $Dist(,)$ is defined as euclidean distance between two points.

Based on the topological structure of \textit{B(k)}, these generated points can be connected to surfaces to form the anchor mesh. However, using this simple anchor mesh can result in significant distortion due to inaccurate selection of corresponding points caused by movements, as demonstrated in Fig.~\ref{simple matched mesh}. To mitigate this issue, we employ motion estimation techniques. Specifically, the motion vector for the current vertex is estimated on the basis of the displacement vectors of neighboring vertices, calculated as follows:
\begin{equation}
m_{est}^{i} = \frac{1}{n} \sum_{l\in N(V_{k}^{i})}^{} m_{}^{l}
\end{equation}
where $N(V_{k}^{i})$ is the set of neighboring vertices around $V_{k}^{i}$ in \textit{B}(k) whose motion vectors have been calculated. $n$ is the quantity of $N(V_{k}^{i})$ and $m_{}^{l}$ represents the motion vector of each neighboring vertex. An offset point is then calculated by applying the estimated motion vector to the vertex. Subsequently, the coarse anchor vertex is determined utilizing the octree-based nearest neighbor search algorithm within \textit{M(t)} based on the offset point. Afterwards, the true motion vector is obtained from the difference between the coarse anchor point $V_{o}^{i}$ and the corresponding vertex in \textit{B(k)}:
\begin{equation}
m_{}^{i} = V_{o}^{i} - V_{k}^{i} \label{motion}
\end{equation}

Although the coarse anchor mesh shares the same topology as \textit{B(t)}, the anchor point must be the vertex of \textit{M(t)}. However, it is not necessarily the closest point in \textit{M(t)} corresponding to the vertex of \textit{B}(k), resulting in a suboptimal solution to (\ref{min}) and poor detail quality of the mesh as shown in Fig.~\ref{4.1}.

\setlength{\belowcaptionskip}{-1pt}
\begin{figure*}[t]
\centering
\begin{subfigure}[b]{0.22\linewidth}
    \centering
    \includegraphics[width=\linewidth]{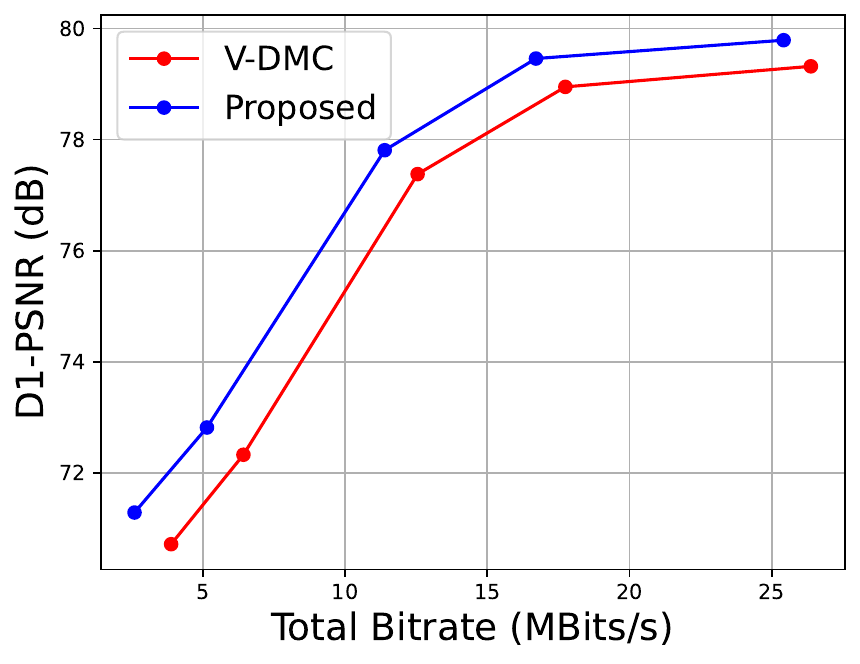}
    \caption{Basketball\_player}
    \label{bask}
\end{subfigure}
\begin{subfigure}[b]{0.22\linewidth}
    \centering
    \includegraphics[width=\linewidth]{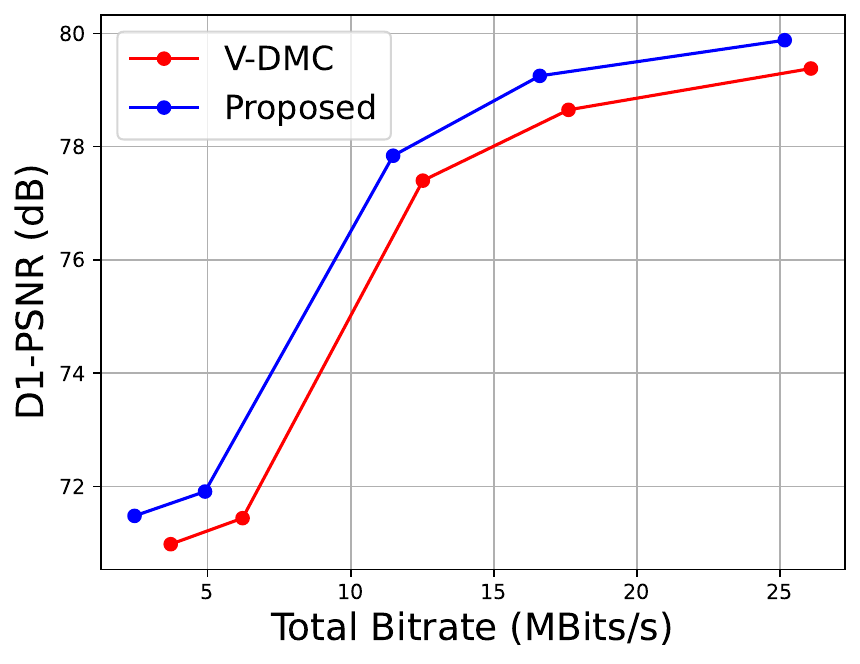}
    \caption{Dancer}
    \label{dancer}
    \end{subfigure}
\begin{subfigure}[b]{0.22\linewidth}
    \centering
    \includegraphics[width=\linewidth]{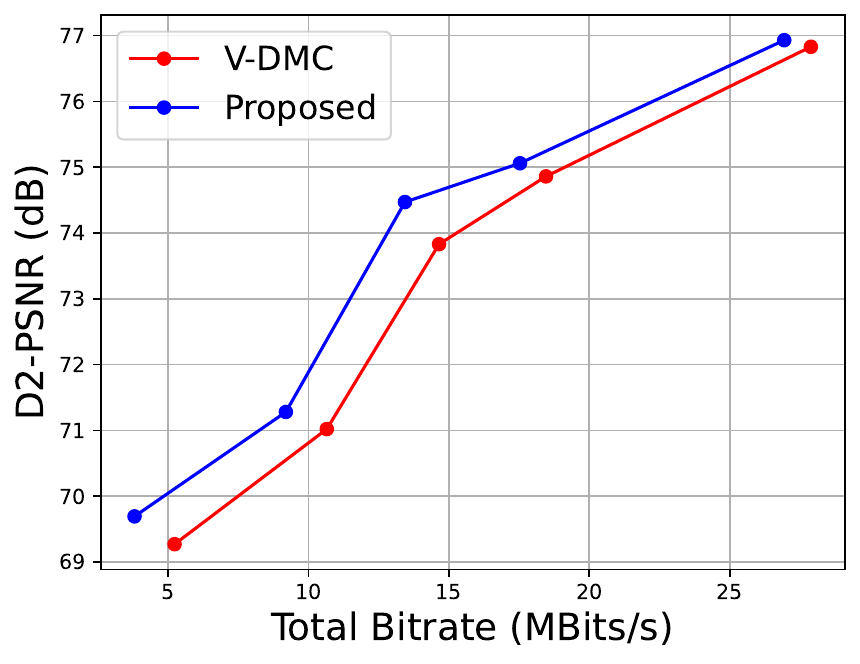}
    \caption{Longdress}
    \label{longdress}
    \end{subfigure}
\begin{subfigure}[b]{0.22\linewidth}
    \centering
    \includegraphics[width=\linewidth]{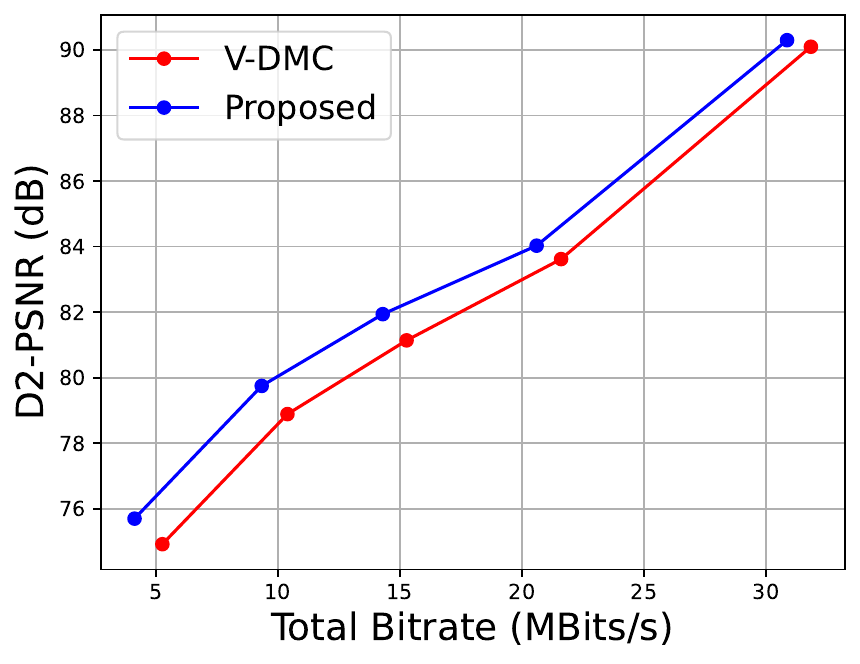}
    \caption{Football\_player}
    \label{Footballplayer}
    \end{subfigure}
\caption{Objective RD performance evaluation.}
\label{Objective RD performance evaluation}
\vspace{-10pt}
\end{figure*}

\subsection{Fine-Stage Anchor Mesh Generation}
To enhance the detail quality, we propose the utilization of the Quadratic Error Metric (QEM) algorithm \cite{garland1997surface} to identify more precise anchor points based on the coarse anchor mesh. This algorithm is utilized for mesh simplification by diminishing the quantity of triangles through the amalgamation of edges with minimal quadratic error into novel vertices. Consequently, the simplified mesh preserves the original shape while maintaining high quality. Therefore, we employ this method to optimize the coarse mesh fidelity by collapsing edges in \textit{M(t)} based on their error calculations, thus creating locally optimal vertices on the mesh surface. The vertex error is delineated as follows:
\begin{align}
\nabla (V) &= \sum_{P\in pl(V)}^{} ((P^{T}V)^{2}) &= V^{T}(\sum_{P\in pl(V)}^{}PP^{T})V
\end{align}
where $\nabla (V)$ represents the error at point $V$ and $pl(V)$ is the set of planes associated with  point $V$. Each $P$ is defined by:
\begin{align}
P = \begin{bmatrix}a & b & c & d\end{bmatrix}^{T}\label{plane}
\end{align}
with coefficients $a$,$b$,$c$ and $d$ satisfying the plane equation $ax + by + cz +d= 0$ and $a^{2} + b^{2} + c^{2} + d^{2} = 1$. The error of an edge $E$, where the point $V$ is located, is expressed as:
\begin{equation}
Q_{E} = q_{V}+q_{V'}
\end{equation}
where $Q_{E}$ is the error matrix for edge $E$. $q_{V}$ is the error matrix of the point $V$, calculated as $\sum_{P\in pl(V)}^{}PP^{T}$. $q_{V'}$ is the error matrix for the other vertex on the edge $E$. Consequently, the solution of the post-collapse point is transformed into an optimization problem:
\begin{equation}
\mathrm{arg} \min_{\overline{V}} \overline{V}_{}^{T}Q_{E}\overline{V}\label{QEM}
\end{equation}
where $\overline{V}$ is the fine anchor point. The collapse-selected edge is determined by identifying the edge with the minimal error among all connected edges to the vertex, optimizing the mesh structure for improved quality and fidelity. The QEM algorithm within \textit{M(t)} is used to compute a fine vertex for each coarse anchor point, iteratively. After each collapse, the error matrices of all affected edges are updated to reflect the new geometric configurations. This iterative process ultimately results in a fine anchor mesh that significantly enhances the quality of the reconstruction.

\subsection{Adaptive Displacements Quantization}
In the current V-DMC, the fine anchor mesh is subdivided to obtain the subdivided mesh. Displacements which are calculated as the differences between the input mesh and the subdivided mesh are quantized using a fixed quantization parameter and encoded by a video codec. However, this method does not consider the importance of individual points, leading to uniform quantization across the mesh. To improve upon this, we propose an adaptive quantization method for displacements based on the number of neighboring points, which can be formulated as:
\begin{align}
D_{q} & = D*\alpha*A + \delta \\
A & = N(V_{D})/\hbar
\end{align}
where $D$ and $D_{q}$ are displacements before and after quantization, respectively. $\delta$ and $\alpha$ are preset displacement offsets and quantization step sizes, both constants. $A$ denotes adaptive weights influenced by the number of neighbors $N(V_{D})$. The hyper-parameter $\hbar$ controls the sensitivity of the weighting to neighbor counts.A point with more neighbors indicates richer local detail, requiring finer displacement quantization. As a result, more bit streams are allocated to these areas to achieve better BD-Rate gain.

\section{Results and Analysis}
\subsection{Experimental Setting}
We evaluate the effectiveness of our method on four dynamic mesh sequences used by MPEG which are challenging for applying inter-frame coding, including Longdress, Basketball\_player, Dancer and Football\_player, each sequence contains over 300 frames. We conduct comparisons over 300 frames against the V-DMC test model version 7.0\cite{TMM}. Five bitrate conditions(R1$\sim$R5) defined by V-DMC CTC\cite{cfp} under Random Access (RA) configuration are tested. 
Additionally, the V-DMC utilizes the point-to-point error (D1) and point-to-plane error (D2) as the object geometry evaluation metrics while utilizing the PSNR of Luma, Chroma Cb, and Chroma Cr as texture evaluation metrics. In fairness, the Bjøntegaarddelta (BD) rate\cite{bjontegaard2001calculation} is used to indicate the average bitrate reduction between the two methods for the same mesh reconstruction.

\afterpage{
\begin{table}[thbp]
\centering
\caption{BD-Rate compared to V-DMC.}
\renewcommand{\arraystretch}{1.4} % 增加行高
\label{table of bdrate}
\begin{tabular}{l|c c c c c}
\hline
                  & D1      & D2      & Luma   & Cb     & Cr     \\ \hline
Basketball\_player & -8.1\%  & -8.1\%  & -0.8\%  & -0.3\%  & -0.3\%  \\ 
Dancer            & -7.2\%  & -7.3\%  & -0.1\% & 0.3\%  & 0.4\%  \\ 
Longdress         & -10.3\% & -10.3\% & -1.7\% & -0.1\% & -0.1\% \\ 
Football\_player   & -7.4\%  & -7.8\%  & -0.1\% & 0.6\%  & 0.4\%  \\ \hline
Encoding time & \multicolumn{5}{c}{\centering 108\%} \\
Decoding time & \multicolumn{5}{c}{\centering 102\%} \\ \hline
% \vspace{-2pt}
\end{tabular}
\vspace{-20pt}
\end{table}

\begin{figure}[thbp]
\centering
\begin{subfigure}[b]{0.25\linewidth}
    \centering
    \includegraphics[width=0.7\columnwidth]{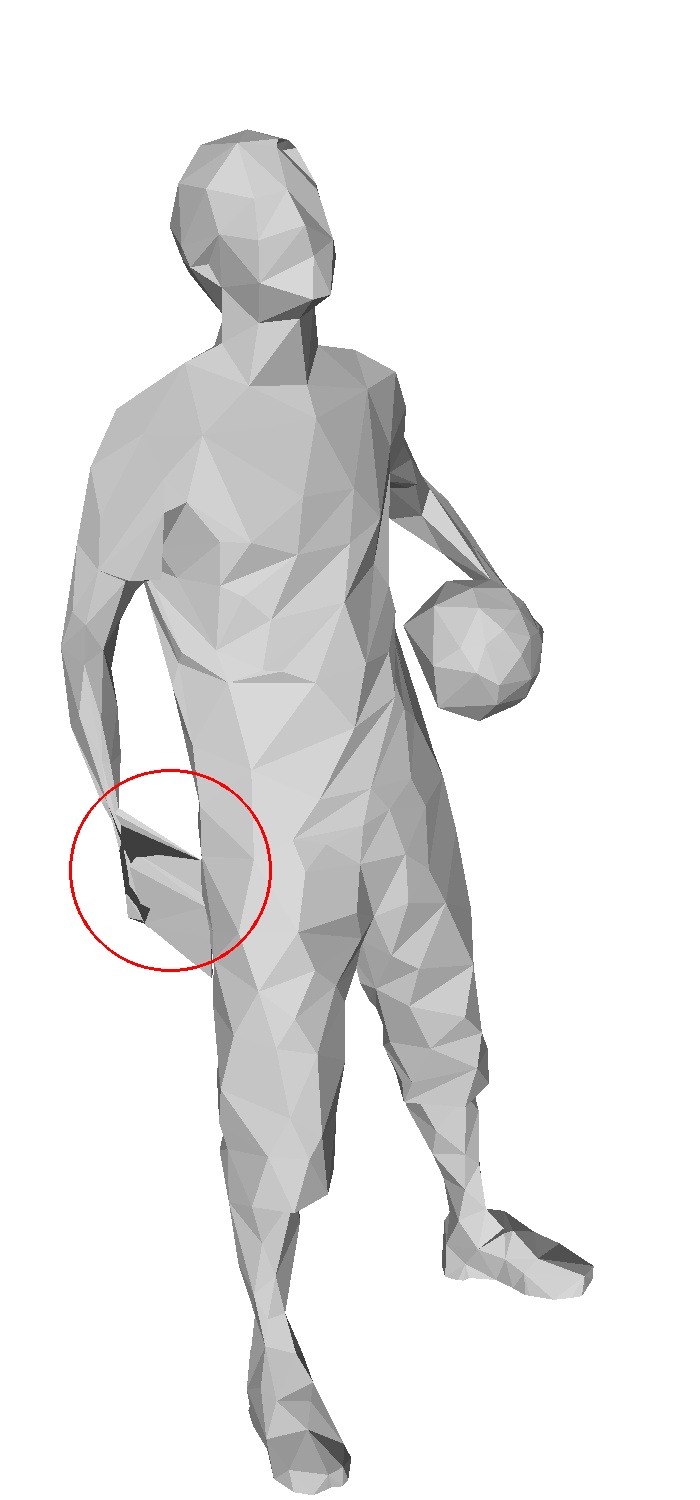}
    \caption{}
    \label{simple matched mesh}
    \end{subfigure}
\begin{subfigure}[b]{0.25\linewidth}
    \centering
    \includegraphics[width=0.7\columnwidth]{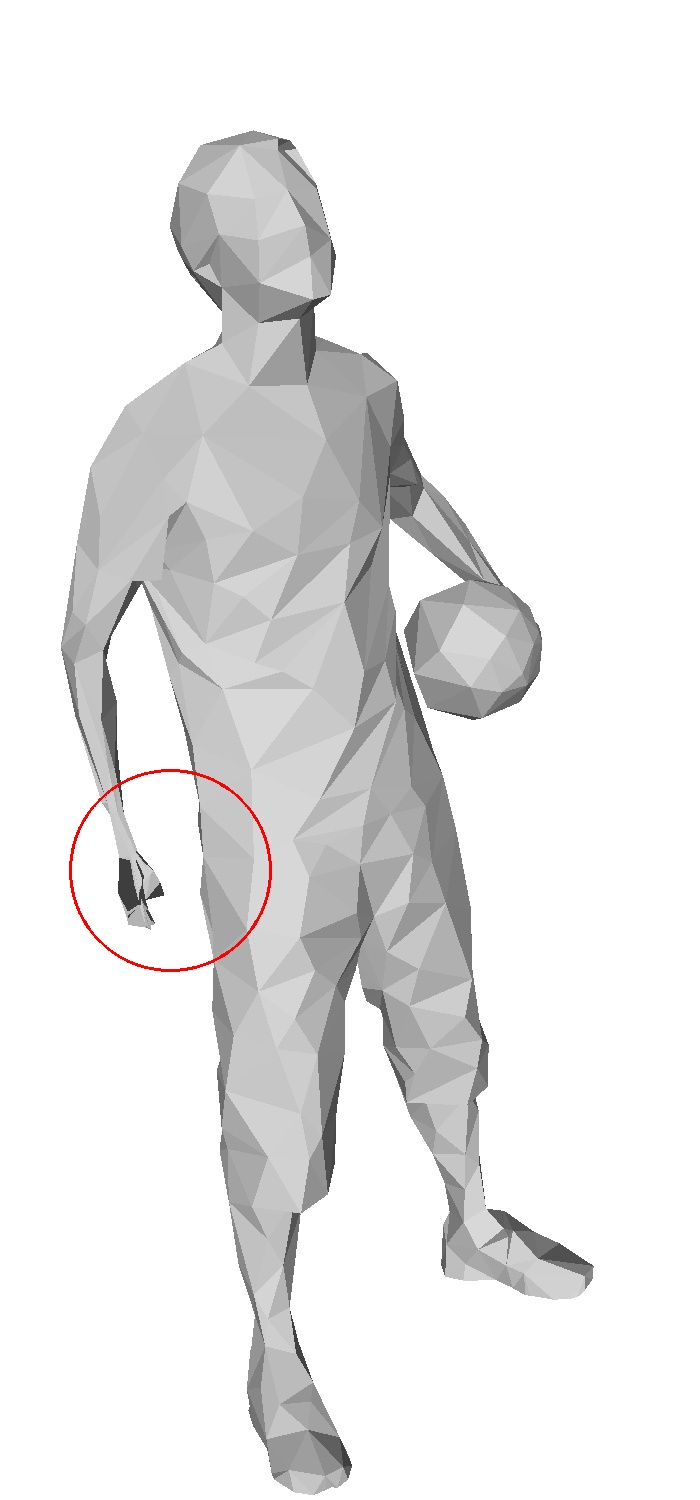}
    \caption{}
    \label{4.1}
    \end{subfigure}
\begin{subfigure}[b]{0.25\linewidth}
    \centering
    \includegraphics[width=0.7\columnwidth]{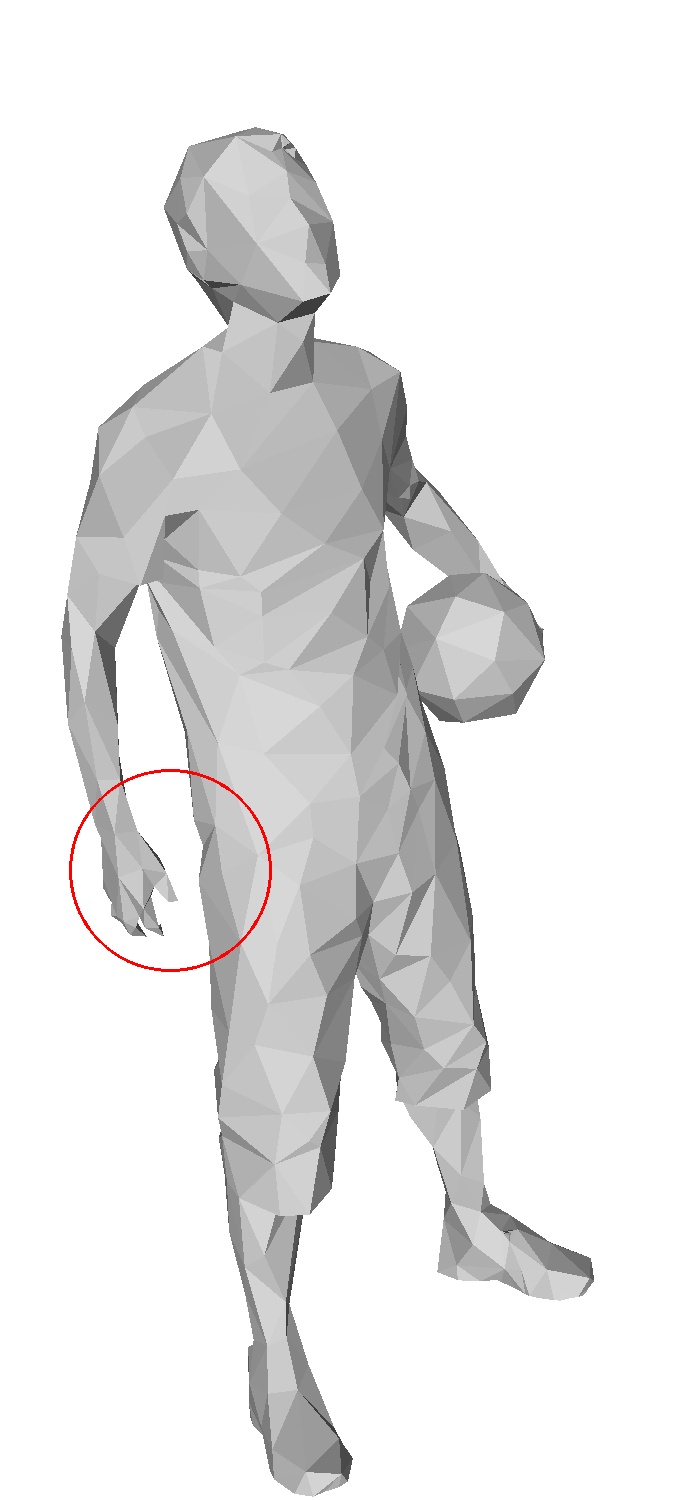}
    \caption{}
    \label{4.2}
    \end{subfigure}
\caption{From left to right: (a) initial anchor mesh generated by octree-based nearest neighbor search, (b) coarse anchor mesh enhanced with motion estimation, and (c) fine anchor mesh optimized with QEM.}
\label{Comparison of matched mesh}
\vspace{0pt}
\end{figure}
}
\subsection{Quantitative Results}
Tab.~\ref{table of bdrate} illustrates the BD-Rate improvements achieved by our method compared to the standard V-DMC. Notably, the D1 and D2 BD-Rate gains for our method are substantial, reaching 7.2\% \( \sim \) 10.3\% for the four sequences. This enhancement is principally attributed to the increased exploitation of inter-frame coding for meshes with dynamic topologies. The progression is facilitated by our innovative coarse-to-fine anchor mesh generation approach, which effectively conserves data while preserving reconstruction fidelity. Moreover, the adaptive quantization of displacements optimizes bitrate allocation, facilitating a more judicious distribution of resources towards regions of heightened significance. Attribute components demonstrate a modest BD-Rate gain in the Longdress and Basketball\_player sequences, albeit with some degradation observed in other sequences. Fig.~\ref{Objective RD performance evaluation} delineates the D1 and D2 rate-distortion curves across five distinct bitrate configurations (R1 to R5). Our methodology achieves bitrate reductions of 10\% to 25\% at lower bitrate levels (R1 to R3) without compromising geometry PSNR. However, at higher bitrates (R4 to R5), the observed savings diminish to a negligible extent. The diminished efficacy at high bitrates can be attributed to the proliferation of vertices, which causes overlap among triangular patches within the components. As a result, the matching mesh's quality is reduced, leading to fewer frames suitable for inter-frame coding and increased temporal redundancy. In terms of computational complexity, our proposed method increases the encoding time to 108\% and the decoding time to 102\%, relative to the original V-DMC benchmarks of 100\% for both encoding and decoding.

\subsection{Qualitative Results}
We conducted a comparative analysis of the anchor mesh between different processes, as illustrated in Fig.~\ref{Comparison of matched mesh}. The initial anchor mesh demonstrated substantial deformations in regions characterized by extensive movements, such as the hands. Following the integration of motion estimation techniques, these deformations were markedly diminished, albeit the quality of detail remained below optimal standards. Ultimately, the implementation of the Quadratic Error Metrics (QEM) method markedly enhanced the perceived quality of the finely anchor mesh, especially in areas necessitating high levels of detail.

\subsection{Ablation Study}
We conducted an ablation study to demonstrate the benefit of each module in our method. Evaluations were performed on 32 frames within the Basketball\_player sequence across five distinct bitrates. As shown in Tab.~\ref{table of ablation study}, the BD-rate gain for D2 PSNR increases from 0.7\% to 6.5\% after applying QEM-based refinement to generate fine anchor mesh. Ultimately, our method can achieve 8.0\%  BD-rate gain for D2 PSNR. Moreover, attribute components are also improved in the process. The table demonstrates that each module of our method provides specific benefits, with QEM-based refinement being especially advantageous.

\begin{table}[tbp]
\centering
\caption{Ablation study of different stages. NNS, MS, QEM, AQ stand for Nearest neighbor search, Motion estimation, QEM-based refinement, and Adaptive Quantization respectively. The baseline is the V-DMC standard.}
\renewcommand{\arraystretch}{1.4} % 增加行高
\label{table of ablation study}
\begin{tabular}{>{\centering\arraybackslash}p{0.4cm} >{\centering\arraybackslash}p{0.4cm} >{\centering\arraybackslash}p{0.4cm} >{\centering\arraybackslash}p{0.4cm}|c c c c c}
\hline
NNS &MS &QEM &AQ  & D1      & D2      & Luma   & Cb     & Cr     \\ \hline
\checkmark & & &     & -0.7\%  & -0.7\%  & 0.9\%  & 1.8\%  & 1.1\%  \\ 
\checkmark &\checkmark & &          & -2.8\%  & -2.1\%  & 0.3\% & 1.1\%  & 0.8\%  \\ 
\checkmark &\checkmark &\checkmark &        & -5.9\% & -6.5\% & -0.5\% & -0.1\% & -0.1\% \\ 
\checkmark &\checkmark &\checkmark &\checkmark   & -7.7\%  & -8.0\%  & -0.6\% & -0.3\%  & -0.3\%  \\ \hline
\end{tabular}
\vspace{-16pt}
\end{table}
\section{Conclusion}
In this paper, we proposed a coarse-to-fine method for generating the anchor mesh, which serves as an effective tool for exploiting temporal correlations within the V-DMC framework. This method ensures that the anchor mesh maintains the same connectivity as the reference base mesh, which is crucial for enabling efficient inter-frame coding. Finally, the experimental results show that our method achieves 7.2\% \( \sim \) 10.3\% BD-rate gain. Additionally, the adaptive displacement quantization approach provides a more strategic bitrate allocation. Our future research will further improve the compression quality of the proposed method in the high-bitrate case.

\section*{Acknowledgment}
This paper is supported in part by National Natural Science Foundation of China (62371290, U20A20185), the Fundamental Research Funds for the Central Universities of China, and STCSM under Grant (22DZ2229005). The corresponding author is Yiling Xu(e-mail: yl.xu@sjtu.edu.cn).

% \section*{References}
\afterpage{
\bibliographystyle{IEEEtran}
\bibliography{mesh_compression}

\begin{thebibliography}{10}
\providecommand{\url}[1]{#1}
\csname url@samestyle\endcsname
\providecommand{\newblock}{\relax}
\providecommand{\bibinfo}[2]{#2}
\providecommand{\BIBentrySTDinterwordspacing}{\spaceskip=0pt\relax}
\providecommand{\BIBentryALTinterwordstretchfactor}{4}
\providecommand{\BIBentryALTinterwordspacing}{\spaceskip=\fontdimen2\font plus
\BIBentryALTinterwordstretchfactor\fontdimen3\font minus \fontdimen4\font\relax}
\providecommand{\BIBforeignlanguage}[2]{{%
\expandafter\ifx\csname l@#1\endcsname\relax
\typeout{** WARNING: IEEEtran.bst: No hyphenation pattern has been}%
\typeout{** loaded for the language `#1'. Using the pattern for}%
\typeout{** the default language instead.}%
\else
\language=\csname l@#1\endcsname
\fi
#2}}
\providecommand{\BIBdecl}{\relax}
\BIBdecl

\bibitem{rossignac1999edgebreaker}
J.~Rossignac, ``Edgebreaker: Connectivity compression for triangle meshes,'' \emph{IEEE transactions on visualization and computer graphics}, vol.~5, no.~1, pp. 47--61, 1999.

\bibitem{mamou2009tfan}
K.~Mamou, T.~Zaharia, and F.~Pr{\^e}teux, ``Tfan: A low complexity 3d mesh compression algorithm,'' \emph{Computer Animation and Virtual Worlds}, vol.~20, no. 2-3, pp. 343--354, 2009.

\bibitem{mammou2022ideo}
K.~Mammou, J.~Kim, A.~M. Tourapis, D.~Podborski, and D.~Flynn, ``Video and subdivision based mesh coding,'' in \emph{2022 10th European Workshop on Visual Information Processing (EUVIP)}.\hskip 1em plus 0.5em minus 0.4em\relax IEEE, 2022, pp. 1--6.

\bibitem{zhu2020view}
W.~Zhu, Z.~Ma, Y.~Xu, L.~Li, and Z.~Li, ``View-dependent dynamic point cloud compression,'' \emph{IEEE Transactions on Circuits and Systems for Video Technology}, vol.~31, no.~2, pp. 765--781, 2020.

\bibitem{fan2023multiscale}
T.~Fan, L.~Gao, Y.~Xu, D.~Wang, and Z.~Li, ``Multiscale latent-guided entropy model for lidar point cloud compression,'' \emph{IEEE Transactions on Circuits and Systems for Video Technology}, vol.~33, no.~12, pp. 7857--7869, 2023.

\bibitem{xia2023learning}
S.~Xia, T.~Fan, Y.~Xu, J.-N. Hwang, and Z.~Li, ``Learning dynamic point cloud compression via hierarchical inter-frame block matching,'' in \emph{Proceedings of the 31st ACM International Conference on Multimedia}, 2023, pp. 7993--8003.

\bibitem{habe2004skin}
H.~Habe, Y.~Katsura, and T.~Matsuyama, ``Skin-off: Representation and compression scheme for 3d video,'' in \emph{Picture Coding Symposium}, 2004, pp. 301--306.

\bibitem{projection}
W.~Zou, H.~Huang, A.~Trioux, and F.~Yang, ``An efficient video-based geometry compression system for 3d meshes,'' in \emph{2023 IEEE International Conference on Visual Communications and Image Processing (VCIP)}, 2023, pp. 1--5.

\bibitem{Static}
T.~N. Canh, F.-Y. Chao, C.~Huang, X.~Xu, and S.~Liu, ``Symmetric geometry coding for static meshes,'' in \emph{2023 IEEE International Conference on Visual Communications and Image Processing (VCIP)}, 2023, pp. 1--5.

\bibitem{han2007time}
S.-R. Han, T.~Yamasaki, and K.~Aizawa, ``Time-varying mesh compression using an extended block matching algorithm,'' \emph{IEEE Transactions on Circuits and Systems for Video Technology}, vol.~17, no.~11, pp. 1506--1518, 2007.

\bibitem{yamasaki2010patch}
T.~Yamasaki and K.~Aizawa, ``Patch-based compression for time-varying meshes,'' in \emph{2010 IEEE International conference on image processing}.\hskip 1em plus 0.5em minus 0.4em\relax IEEE, 2010, pp. 3433--3436.

\bibitem{jin2023inter}
X.~Jin, J.~Xu, and K.~Kawamura, ``Inter-frame coding for dynamic meshes via temporally-consistent re-meshing,'' in \emph{2023 IEEE International Conference on Image Processing (ICIP)}.\hskip 1em plus 0.5em minus 0.4em\relax IEEE, 2023, pp. 2000--2004.

\bibitem{octree}
L.~Wei, S.~Wan, X.~Ding, and Z.~Wang, ``Optimization of octree-based adaptive geometry quantization via up-sampling for g-pcc,'' in \emph{2023 IEEE International Conference on Visual Communications and Image Processing (VCIP)}, 2023, pp. 1--5.

\bibitem{garland1997surface}
M.~Garland and P.~S. Heckbert, ``Surface simplification using quadric error metrics,'' in \emph{Proceedings of the 24th annual conference on Computer graphics and interactive techniques}, 1997, pp. 209--216.

\bibitem{TMM}
``V-dmc tmm 7.0,'' \emph{ISO/IEC JTC 1/SC 29/WG 7}, March 2024.

\bibitem{cfp}
``Cfp for dynamic mesh coding,'' \emph{ISO/IECJTC1/SC29/WG7/N00231}, October, 2021.

\bibitem{bjontegaard2001calculation}
G.~Bjontegaard, ``Calculation of average psnr differences between rd-curves,'' \emph{ITU SG16 Doc. VCEG-M33}, 2001.

\end{thebibliography}
}

\end{document}